\documentstyle[epsfig,here]{aipproc}
\begin{document}
\date{  }
\title{
 Heavy Meson Production at COSY~-~11    }
\author{ 
P.~Moskal$^{(a)}$,
H.~-H.~Adam$^{(b)}$,
J.~T.~Balewski$^{(c,d,e)}$,
A.~Budzanowski$^{(c)}$, 
%
C.~Goodman$^{(e)}$,
D.~Grzonka$^{(d)}$,
L.~Jarczyk$^{(a)}$,
M.~Jochmann$^{(f)}$,
A.~Khoukaz$^{(b)}$,
K.~Kilian$^{(d)}$,
P.~Kowina$^{(g)}$,
M.~K\"ohler$^{(f)}$,
T.~Lister$^{(b)}$,
W.~Oelert$^{(d)}$,
C.~Quentmeier$^{(b)}$,
R.~Santo$^{(b)}$,
G.~Schepers$^{(b,d)}$,
U.~Seddik$^{(h)}$,
T.~Sefzick$^{(d)}$,
S.~Sewerin$^{(d)}$,
J.~Smyrski$^{(a)}$,
A.~Strza{\l}kowski$^{(a)}$,
M.~Wolke$^{(d)}$,
P.~W{\"u}stner$^{(f)}$
   }
\address{
$^{(a)}$ Institute of Physics, Jagellonian University,
PL--30059 Cracow, Poland \\
$^{(b)}$ IKP, Westf. Wilhelms-Universit\"at, Wilhelm-Klemm-Stra{\ss}e 9,
D--48149 M{\"u}nster, Germany \\
$^{(c)}$ Institute of Nuclear Physics, ul. Radzikowskiego 152, PL-31-342 Cracow, Poland \\
$^{(d)}$ Institut f\"ur Kernphysik, Forschungszentrum J\"ulich,
D--52425 J{\"u}lich, Germany\\
$^{(e)}$ IUCF, Milo B. Samson Lane, Bloomington, IN 47405, USA \\
$^{(f)}$ ZEL, Forschungszentrum J{\"u}lich, D--52425 J{\"u}lich, Germany \\
$^{(g)}$ Institute of Physics, Silesian University, PL-40-007 Katowice, Poland \\
$^{(h)}$ Egyptian Atomic Energy Authority, 101 Sharia Kaser El-Aini, 13759 Cairo, Egypt \\
  }
        \maketitle
\begin{abstract}
   The COSY-11 collaboration has measured the total cross section for the
  $pp\rightarrow pp\eta^{\prime}$ and $pp\rightarrow pp \eta$ reactions
  in the excess energy range from
  Q~=~1.5~MeV to Q~=~23.6~MeV and  from Q~=~0.5~MeV to Q~=~5.4~MeV, respectively.
  Measurements have been performed with the  total luminosity
  of 73~nb$^{-1}$ for the  $pp \rightarrow pp\eta$ reaction
  and  1360~nb$^{-1}$ for the $pp \rightarrow pp \eta^{\prime}$ one.
  Recent results are presented and discussed.


\end{abstract}

%
%
%

%
   \renewcommand{\topfraction}{0.90}

\section*{Introduction}
 The word heavy used in the title requires a short explanation.
The reason is rather historical, and it seems now that heavy are
all mesons but not pions. The talk will concern the production of the
$\eta$ and $\eta^{\prime}$ mesons, and since
 $\eta^{\prime}$ is even heavier than $\eta$ the discussion concerning this meson
will constitute the major part of the presentation.

     Last year, for the first time total cross
sections for the production of the $\eta^{\prime}$ meson
in the collision of protons  
close to the reaction threshold
have been published~\cite{hiboupl,moskalprl}.
Two independent experiments performed at the accelerators
SATURNE and COSY have delivered consistent results.

  The first remarkable inference derived from  these experiments was that the 
total cross sections for the $pp \rightarrow pp\eta^{\prime}$ reaction
are  by about a factor of fifty smaller than the cross sections for the $pp \rightarrow pp\eta$
reaction at the corresponding values of the excess energy. 
Trying to explain this large difference Hibou et al.~\cite{hiboupl}
showed that the one-pion-exchange model with the parameters adjusted to fit
the total cross section for the $pp \rightarrow pp\eta$ reaction
underestimate the $\eta^{\prime}$ data by about a factor of two. 
 This discrepancy suggests that short-range production mechanisms
as for example heavy meson exchange, 
mesonic currents~\cite{nakayama}, or more exotic processes like the production
via a fusion of  gluons~\cite{bass} may contribute significantly in the
creation of $\eta$ and $\eta^{\prime}$ mesons~\cite{wilk}. Especially that the
momentum transfer required to create these mesons is much larger 
compared to the pion production, and already in case of pions a significant contribution
from the short-range heavy meson exchange is necessary in order to obtain
agreement with the experiments~\cite{haiden,horowitz}.

  The second interesting observation was that 
the energy dependence of the total cross section for the $pp \rightarrow pp\eta$ and
$pp \rightarrow pp\eta^{\prime}$  reactions 
does not follow the predictions based on the phase space volume and the proton-proton
final state interaction, which is the case in the $\pi^{0}$ meson production~\cite{meyer90,meyer92}.
   Moreover, for $\eta$ and $\eta^{\prime}$ mesons 
the deviation from this prediction were qualitatively different.
Namely, the close to threshold cross sections for the $\eta$ meson
are strongly enhanced compared to the model comprising only the proton-proton interaction~\cite{caleneta} 
in contrary to the observed suppression in the case of the meson $\eta^{\prime}$. 
  The energy dependence of the total cross section for the $pp \rightarrow pp\eta$ reaction
can be, however, explained  when the $\eta$-proton attractive interaction is taken into account~\cite{ulf95,moalem1}.
Albeit $\eta$-proton interaction is much weaker than the proton-proton one 
(compare scattering length a$_{p\eta} = 0.751$~fm + $i$~0.274~fm~\cite{greenwycech}  with a$_{pp} = -7.83$~fm~\cite{naisse})
it becomes important through the interference terms between the various final pair interactions~\cite{moalem1}.
  By analogy, the steep decrease of the total cross section when approaching a kinematical threshold
for the $pp \rightarrow pp\eta^{\prime}$ reaction
could have been explained assuming a repulsive 
$\eta^{\prime}$-proton interaction~\cite{moskalacta,baruej}.
This interpretation, however, should rather be excluded
now in view of the new COSY-11 data which will be presented in the next chapters.

%
%
%
%
%
%
%
  
\section*{Possible production mechanisms}

\begin{figure}[H]
\centerline{\epsfig{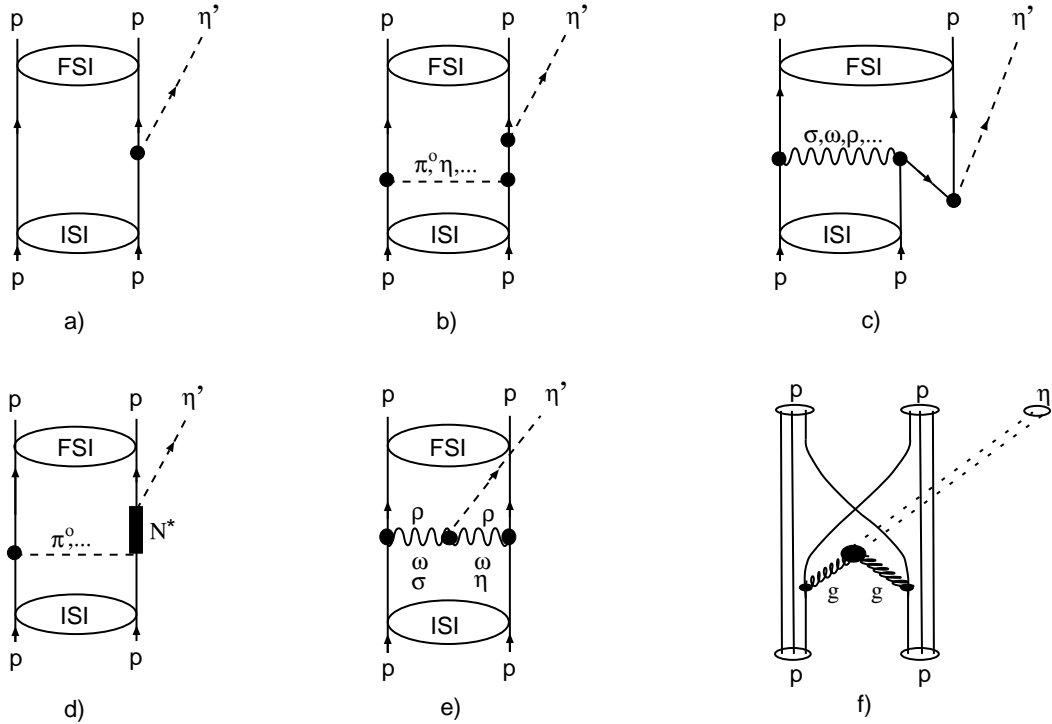}}
\caption{ Diagrams for the $pp \rightarrow pp\eta^{\prime}$ reaction near-threshold: 
          (a)--- $\eta^{\prime}$-bremsstrahlung (nucleonic current)
          (b)--- ``rescattering'' term (nucleonic current)
          (c)--- production through the heavy meson exchange 
          (d)--- excitation of an intermediate resonance (nucleon resonance current)
          (e)--- emission from the virtual meson (mesonic current) 
          (f)--- production via a fusion of gluons (gluonic current).
         }
\label{grafy}
\end{figure}

  The theoretical studies  of the mechanisms accounting for the
$\pi^{0}$ and $\eta$ mesons creation in the close to threshold $pp\rightarrow pp\pi^{0}(\eta)$ reactions
have shown that the short-range component of the N-N force and the off-shell
pion rescattering dominate the production process of the $\pi^{0}$  meson~\cite{haiden,horowitz,hernandez},
whereas the $\eta$ meson  is predominantly produced through the excitation of the intermediate
baryonic resonance~\cite{faldtwilk1,germondwilk,laget,moalem,caleneta}. 
However, the comparison of the experimentally determined
$\eta$ and $\eta^{\prime}$
total cross section ratio with the predictions based on the one-pion-exchange model indicates that 
we are still far from the full understanding of the dynamics of the discussed processes.
In particular, at present there is not much  known about the relative contribution of the possible
reaction mechanisms to the production of the meson $\eta^{\prime}$.
It is expected that similarly as in the case of pions the $\eta^{\prime}$ meson can be produced as depicted 
in Figures~\ref{grafy}a,b,c,d. 
   However, because of the much larger four-momentum transfer, 
short-range mechanisms, like heavy meson exchange (Figure~\ref{grafy}c) or depicted in Figure~\ref{grafy}e
production via a mesonic current,
 where the $\eta^{\prime}$ is created in a fusion of 
 exchanged virtual $\omega$, $\rho$,
or $\sigma$ mesons shell contribute even more significantly.
Recently Nakayama et al.~\cite{nakayama}, studied contributions from the 
 {\em nucleonic }~(Fig.~\ref{grafy}b),
 {\em nucleon resonance }~(Fig.~\ref{grafy}d), and
 {\em mesonic }~(Fig.~\ref{grafy}e) currents and found that each
one separately could describe the absolute values and energy dependence of the 
close to threshold $\eta^{\prime}$ data points~\cite{hiboupl,moskalprl}
after an appropriate adjustment of the ratio of the pseudoscalar to the
pseudovector  coupling. This rather pessimistic conclusion means that  it is  not possible to judge 
about the mechanisms responsible for the $\eta^{\prime}$ meson production
from the total cross section alone.
 Moreover, the possible gluonium admixture in the meson $\eta^{\prime}$ makes the study
even more complicated but certainly also more interesting. 
Figure~\ref{grafy}f depicts appropriate short-range mechanism 
which may lead to the creation of the 
flavour singlet state via a fusion of gluons emitted from the exchanged
quarks of the colliding protons~\cite{kolacosynews}. 
Albeit the quark content of $\eta$ and $\eta^{\prime}$ mesons is very similar,
this manner of the production should contribute primarily
in the creation of the meson $\eta^{\prime}$.
 This is  due to the small pseudoscalar mixing angle
($\Theta_{PS} \approx -15^{o}$)~\protect\cite{bram97}
which implies that the $\eta^{\prime}$ meson is predominantly a flavour singlet state
and is expected to contain a significant admixture of gluons. Further,
it is almost two times  heavier than $\eta$ and hence its creation
requires much larger momentum transfer which is  more probable to be realized
in the short-range interactions.
Unfortunately, at present there are no theoretical calculations
concerning this mechanism.
Now, since the effective coupling constant describing the $\eta^{\prime}$-proton-proton vertex is not known,
 it is even  not possible to determine the contribution from the simplest possible
production mechanism where the $\eta^{\prime}$ is supposed to be emitted as a bremsstrahlung radiation
from one of the colliding protons as it is shown in Figure~\ref{grafy}a.  
Therefore, investigations of the $\eta^{\prime}$ production have to deal with a few problems at the 
same time. Namely: unknown reaction mechanism, unknown coupling constant, and unknown 
proton-$\eta^{\prime}$ interaction. 
 In the next section the present status of the knowledge about the effective
NN$\eta^{\prime}$ coupling constant will be given.
        

\subsection*{$NN\eta^{\prime}$ coupling constant}
   \label{couplconstsec}
     In the effective Lagrangian approach~\cite{zhan95,mukh96}
   the strength of the nucleon-$\eta^{\prime}$
   coupling is driven by the  the $NN\eta^{\prime}$ coupling constant $g_{NN\eta^{\prime}}$,
   which comprises the information about the structure of the $\eta^{\prime}$ meson and the nucleon.
   The knowledge  of the  coupling constant is necessary in the calculation of the production cross
   section if one considers the Feynman diagrams as illustrated in Figure~\ref{grafy}.

   The main difficulty in the determination of this quantity
   is due to the fact that usually the direct production on the nucleon
   is either associated with the production through  baryonic resonances,
   as in the case of the $\gamma p \rightarrow \eta^{\prime} p$ reaction~\cite{ploetzke},
   or with  the exchange of other mesons.
   Therefore, if the direct production mechanism is not dominant it is not possible to extract
   the $NN\eta^{\prime}$ coupling without the clear understanding of the other mechanisms.
   However, it would be very interesting to determine
   the  $g_{NN\eta^{\prime}}$ coupling constant and to compare it with the  
   calculations performed on the quark level assuming 
   the $\eta^{\prime}$ meson structure.  First theoretical considerations concerning this issue
   have been published last year~\cite{lehmann}.

   Assuming that the $\eta$ and $\eta^{\prime}$ mesons are  mixtures of the SU(3) singlet and octet states,
   one can relate the $NN\eta$ and $NN\eta^{\prime}$ coupling constants by the
   following equation~\cite{zhan95,moskalphd}:
     \begin{equation}
        g_{NN\eta^{\prime}} = \frac{sin\Theta + \sqrt{2}cos\Theta}
                                   {cos\Theta - \sqrt{2}sin\Theta} \cdot g_{NN\eta}
                                  \stackrel{\Theta=-15.5^{\circ}}
                                                   {=\!\!\!=\!\!\!=\!\!\!=\!\!\!=\!\!\!
                                                    =\!\!\!=\!\!\!=\!\!\!=\!\!\!=\!\!\!
                                                    =\!\!\!=\!\!\!=\!\!\!=\!\!\!=\!\!\!
                                                    =\!\!\!=\!\!\!=\!\!\!=} 0.82 \cdot g_{NN\eta}.
        \label{getaetap}
     \end{equation}
   The measurements of the $\gamma p \rightarrow p \eta$~\cite{benm95,benm91} reaction have yielded
    that:
     \begin{math}
       0.2~\leq g_{NN\eta}~\leq~6.2,
     \end{math}
   whereas the comparison of the $\pi^{-}p\rightarrow \eta n$ and $\pi^{-}p\rightarrow \pi^{0} n$ reaction
  cross sections implies~\cite{benm95}:
    \begin{math}
       5.7~\leq~g_{NN\eta}~\leq~9.0
    \end{math}.
    The above inequalities and equation~\ref{getaetap} lead to the following range for the $g_{NN\eta^{\prime}}$ value:
     \begin{math}
      {\bf 4.7~\leq~g_{NN\eta^{\prime}}~\leq~5.1},
      \label{getapvalue}
     \end{math}
    which is to be compared to the $\eta^{\prime}$ coupling determined from the fits to low energy
    nucleon-nucleon scattering in the one-boson-exchange models amounting to $g_{NN\eta^{\prime}}=7.3$~\cite{nagels}.

    On the other hand, the $g_{NN\eta^{\prime}}$ coupling constant determined via dispersion methods~\cite{greinkroll}
    turns out to be smaller than~1, $ g_{NN\eta^{\prime}} < 1 $,
    which is in contradiction to the above estimations.

   The $g_{NN\eta^{\prime}}$ coupling constant is also related
   to the issue of the total quark contribution to the proton spin~($\Delta\Sigma$).
   The approximate equation derived in reference~\cite{efre90} reads:
    \begin{math}
      \Delta\Sigma = \Delta u + \Delta d + \Delta s
       =\frac{\sqrt{3} f_{\eta^{\prime}}}{2M}  g_{NN\eta^{\prime}},
       \label{sigmacoupling}
    \end{math}
  where, $f_{\eta^{\prime}}\approx 166~MeV$~\cite{efre90} denotes
  the $\eta^{\prime}$ decay constant and M  the proton mass.
  $\Delta u$, $\Delta d$ and $\Delta s$ are the contributions from
  up, down and strange quarks, respectively~\footnote{
          Contribution of quarks heavier than the $strange$ quark are usually not considered, but
          I.~Halperin and A.~Zhitnitsky  suggested~\cite{halp97} that the intrinsic charm component
          of proton may also carry a significant amount of the proton spin.
          The quark and gluon contributions to the proton spin are widely discussed
          in the literature~\cite{efre90,SMC,SMC2,SMC3,elli96,shor90,hugh88,ashm88}
          based on  measurements of the  spin asymmetries in deep-inealstic scattering of polarised muons on
          polarised protons.
      }.
The  total contribution of the quarks
  to the proton spin  amounts to $\Delta\Sigma=0.38^{+0.09}_{-0.10}$~\cite{SMC}.
   Applying this value in the above equation
   one obtains $ g_{NN\eta^{\prime}} = 2.48^{+0.59}_{-0.65} $, which is consistent with the 
upper limit \begin{math}(g_{NN\eta^{\prime}}~\leq~2.5)\end{math}
set from the comparison of the measured total cross section values for the $pp \rightarrow pp\eta^{\prime}$
reaction 
with the calculations based on the effective Lagrangian approach, where only a direct production 
has been considered~\cite{moskalprl}.

 The present estimations  for  $g_{NN\eta^{\prime}}$ inferred from  different
experiments  are widely spread from 0.2 to 7.3 and are not consistent with each others. 
Therefore more effort is needed on experimental as well as theoretical  side to fix this
important parameter. 

%

\section*{The COSY~-~11 Experiment} 
\begin{figure}[H]
\centerline{\epsfig{figure=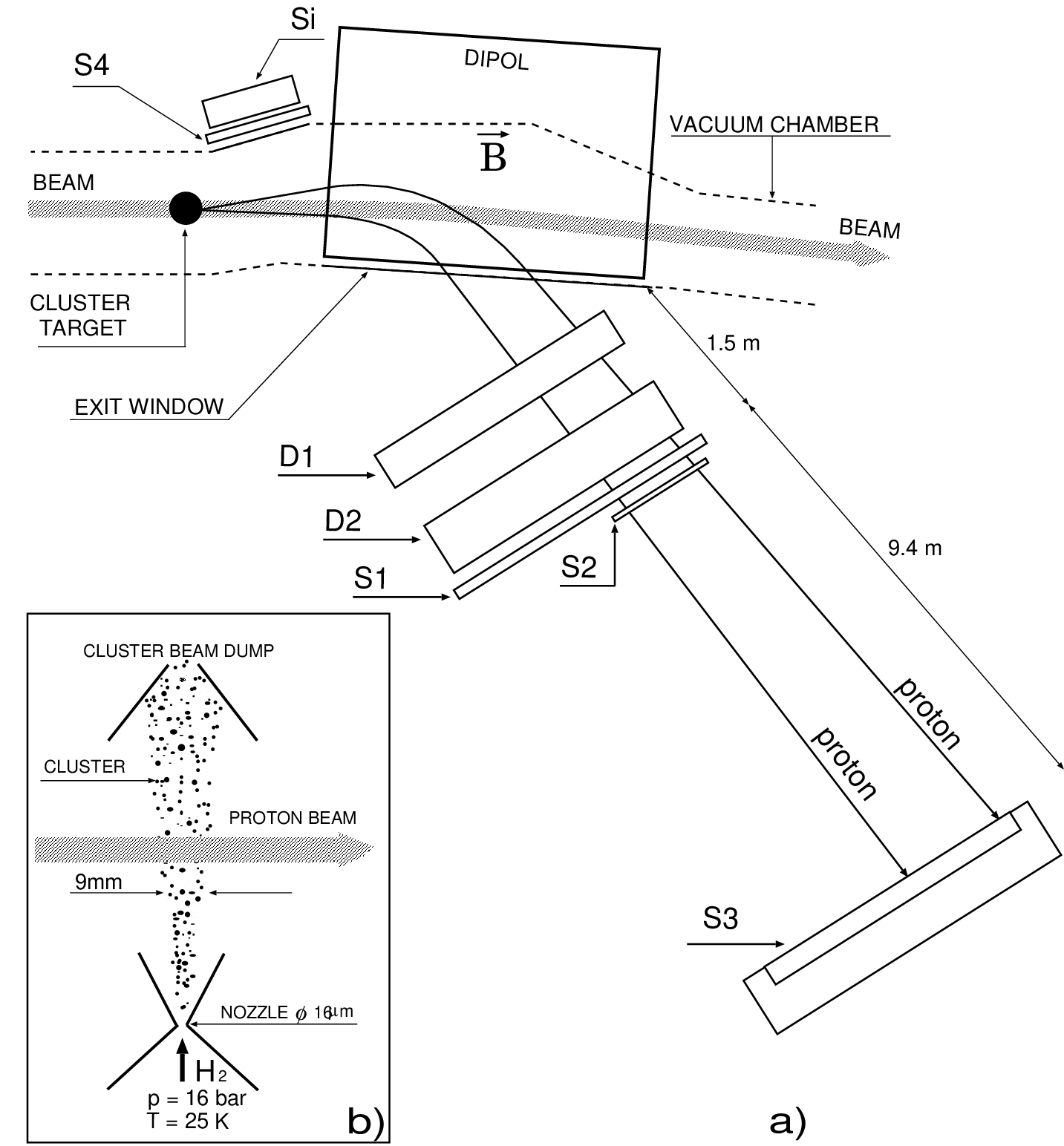,height=13.0cm,angle=0}}
\vspace{0.4cm}
\caption{ a) Schematic view of the COSY-11 detection setup~\protect\cite{brau96}. 
             Only detectors used for the measurements of the $pp\rightarrow pp\eta(\eta^{\prime})$
             reactions are shown. \protect\\
             The cluster target is located in front of the accelerator dipole magnet.
             Protons from
             the $pp\rightarrow pp\eta(\eta^{\prime})$ reaction are bent by the
             magnetic field of the dipole magnet,  whereas
             the beam particles keep circulating in the COSY ring.
             The decay products of the $\eta$ or $\eta^{\prime}$ mesons are not shown, since
          the analysis is based on the measurement of the four-momenta of
          the outgoing protons,
             which  leave the vacuum chamber through the thin exit foil
             and are detected: i)~in two drift chamber stacks D1, D2, ii)~in the scintillator
             hodoscopes S1, S2, iii)~and in the scintillator wall~S3.\protect\\
             For the measurement of the elastically scattered protons, additionally,  scintillation detector S4
             and silicon pad detector Si are used in coincidence with the S1, D1 and D2 detectors.\protect\\
          b) Schematic view of the cluster target.
         }
\label{tarczawiazka}
\end{figure}

 The experiments were performed at the cooler synchrotron COSY-J{\"u}lich~\cite{maie97} which
accelerates protons up to a momentum of $3500~MeV/c$. The threshold momenta for the 
$pp \rightarrow pp\eta$ and $pp \rightarrow pp\eta^{\prime}$
reactions are equal to ${\bf 1981.6~MeV/c}$ and ${\bf 3208.3~MeV/c}$, respectively. 
 About $2\cdot 10^{10}$ accelerated protons
circulate in the ring passing $~1.6\cdot 10^{6}$ times per second through the $H_{2}$ cluster
target~\cite{domb97,khou96} installed in front of one of the dipole magnets,
as depicted schematically in Figure~\ref{tarczawiazka}.
\label{aabc}
The target
  is realized as a beam of $H_{2}$ molecules grouped inside clusters of up to $10^{5}$ atoms.

At the intersection point of the cluster beam with  the COSY proton beam the collision of protons
may result for example in the production of the $\eta^{\prime}$ meson.
 The ejected protons  of the $pp\rightarrow pp \eta(\eta^{\prime})$ reaction,
  having smaller momenta than the beam protons, are separated
 from the circulating beam by the magnetic field. 
 Further they leave the vacuum chamber through a thin exit foil
 and are registered by the detection system consisting of drift chambers and scintillation
 counters as depicted in~Figure~\ref{tarczawiazka}.

%
%
The measurement of the track direction by means  of the drift chambers,
and the knowledge of both the dipol magnetic field
and  the target position allow to reconstruct the momentum vector for each registered particle.
The time of flight measured between the S1 (S2) - and the S3 scintillators gives the particle velocity.
Having momentum and velocity for each particle one can calculate its mass, and hence
identify it.

\begin{figure}[H]
\vspace{-0.5cm}
\centerline{\epsfig{figure=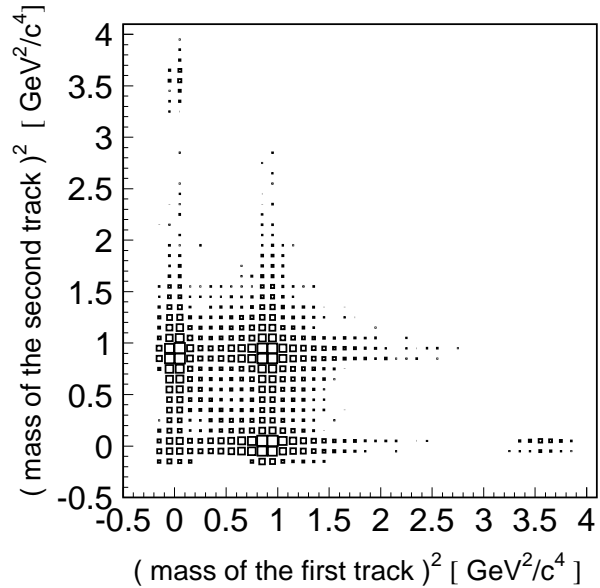,height=9.0cm,angle=0}}
\caption{ Squared masses of two positively charged particles measured in coincidence.
          Pronounced peaks are to be recognized when two protons, proton and pion, two pions, 
          or pion and deuteron were registered. Note that the number of events is 
          shown in logarithmic scale.
         }
\label{invmass}
\end{figure}

\begin{figure}[H]
 \unitlength 1.0cm
  \begin{picture}(12.2,15.0)
    \put(-0.5,0.0){
       \epsfig{figure=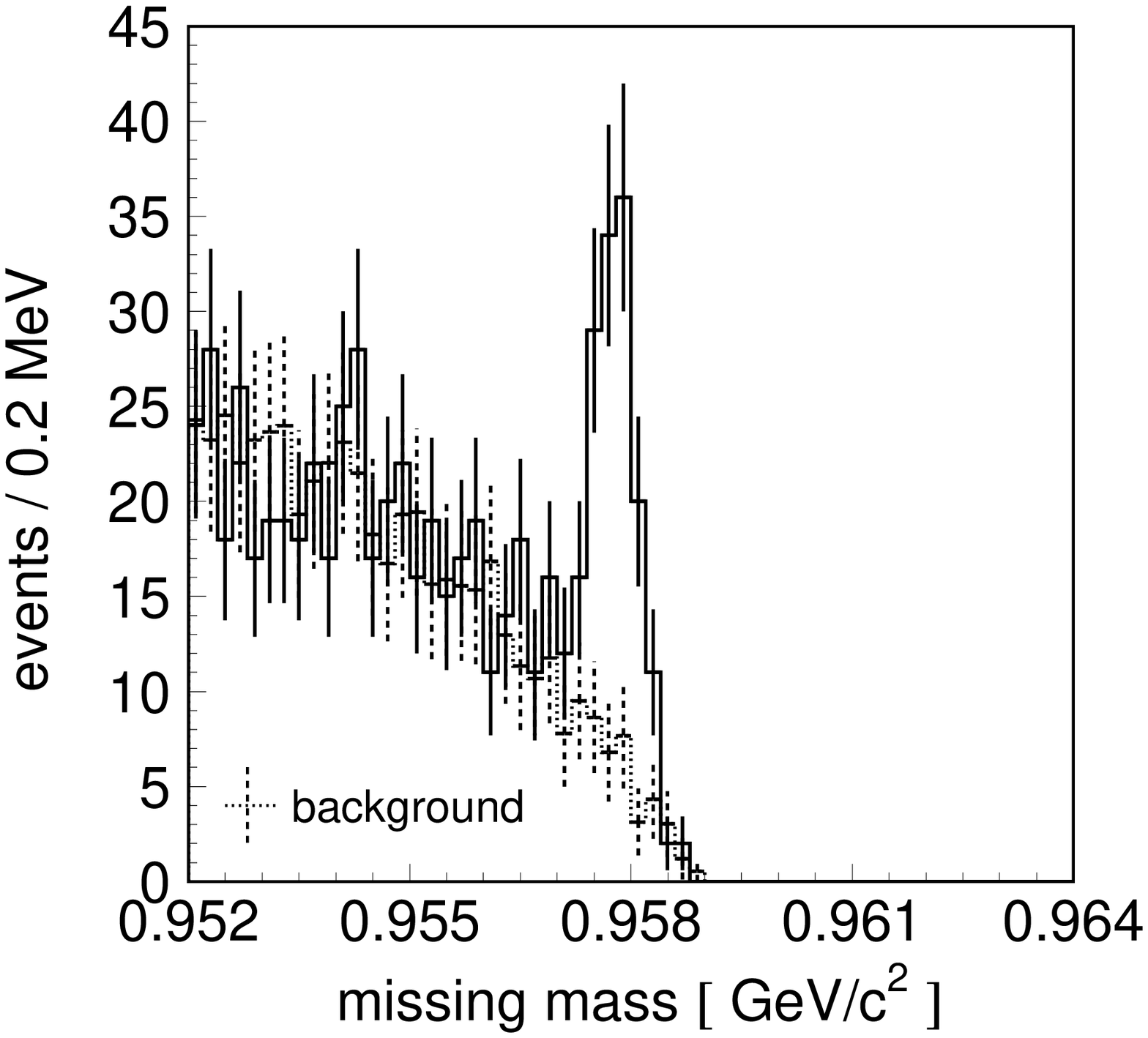,height=6.2cm,width=7.0cm,angle=0}
    }
       \put(5.0,3.7){
          {\large{ e)}}
       }
    \put(7.0,0.0){
       \epsfig{figure=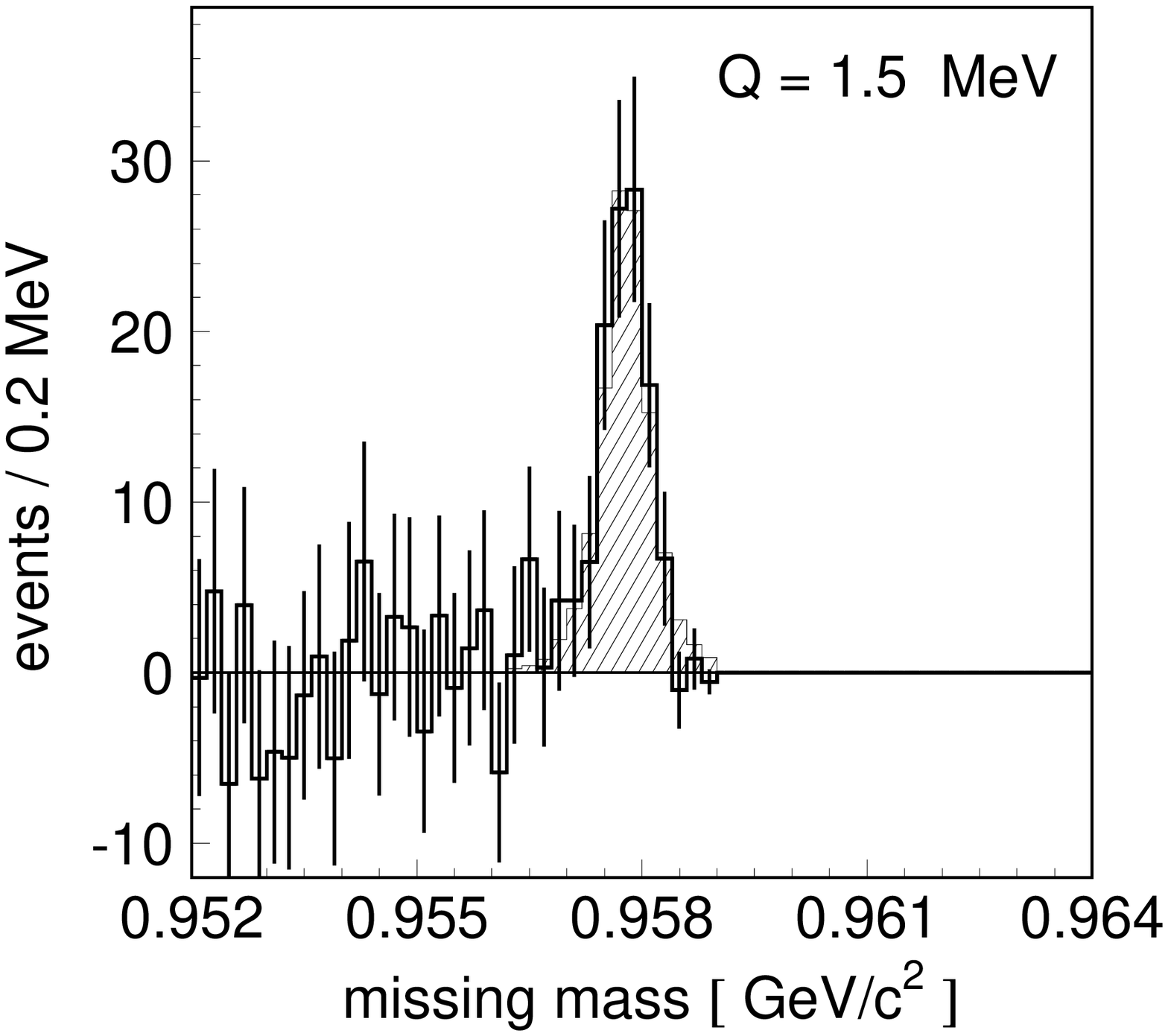,height=6.2cm,width=7.0cm,angle=0}
    }
       \put(12.5,3.7){
          {\large{ f)}}
       }
    \put(-0.5,5.0){
       \epsfig{figure=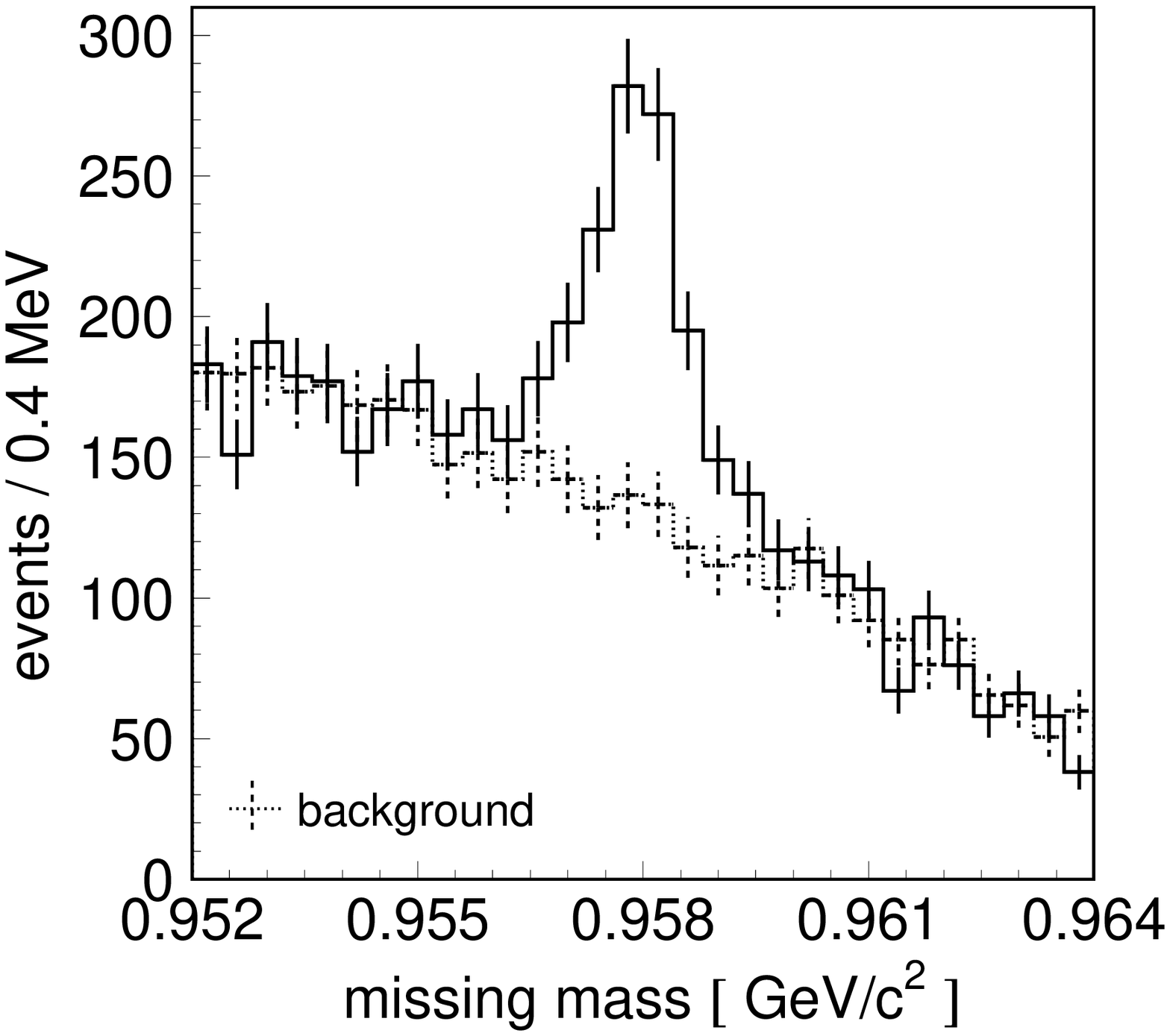,height=6.2cm,width=7.0cm,angle=0}
    }
       \put(5.0,8.7){
          {\large{ c)}}
       }
    \put(7.0,5.0){
       \epsfig{figure=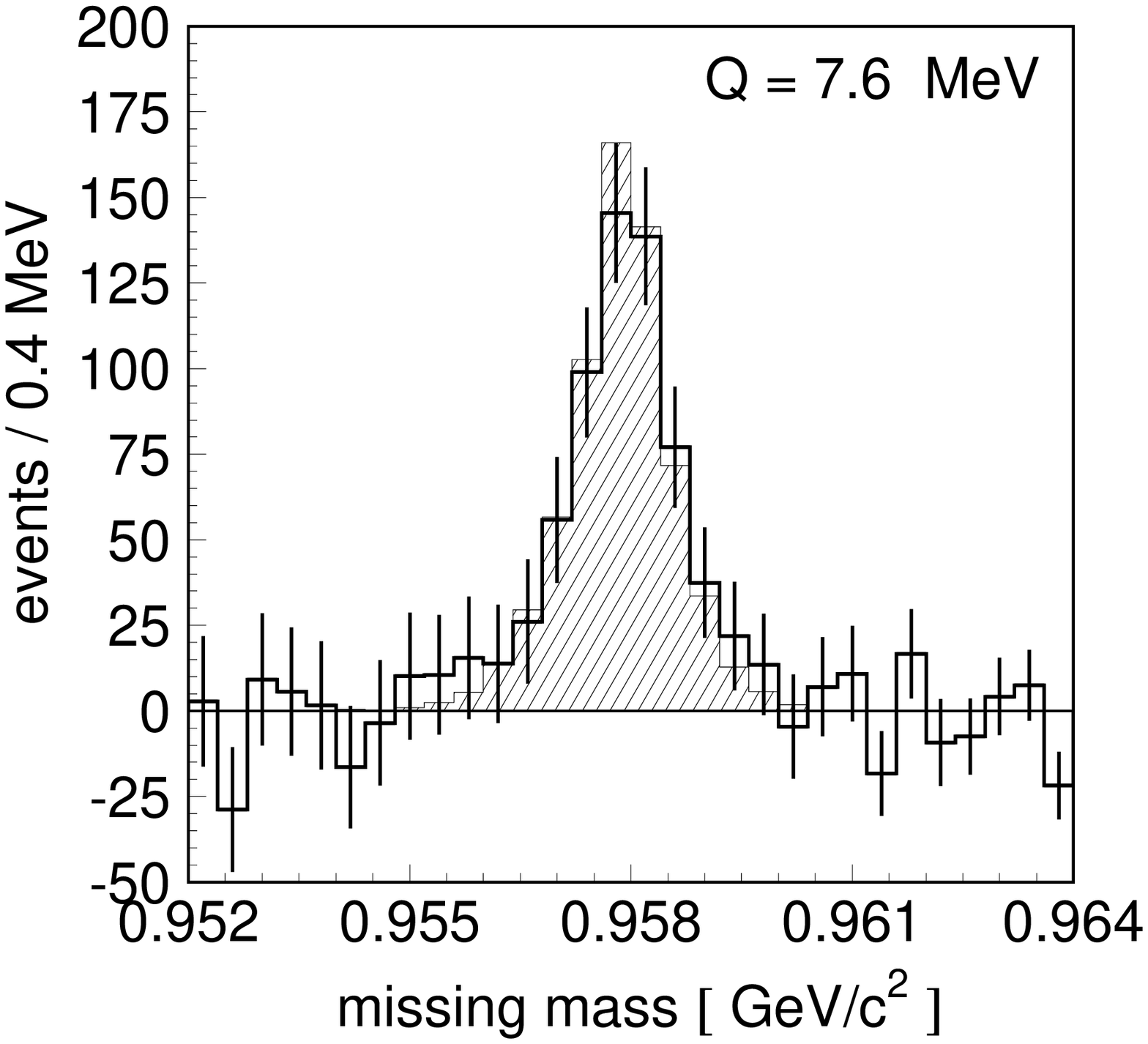,height=6.2cm,width=7.0cm,angle=0}
    }
       \put(12.5,8.7){
          {\large{ d)}}
       }
    \put(-0.5,10.0){
       \epsfig{figure=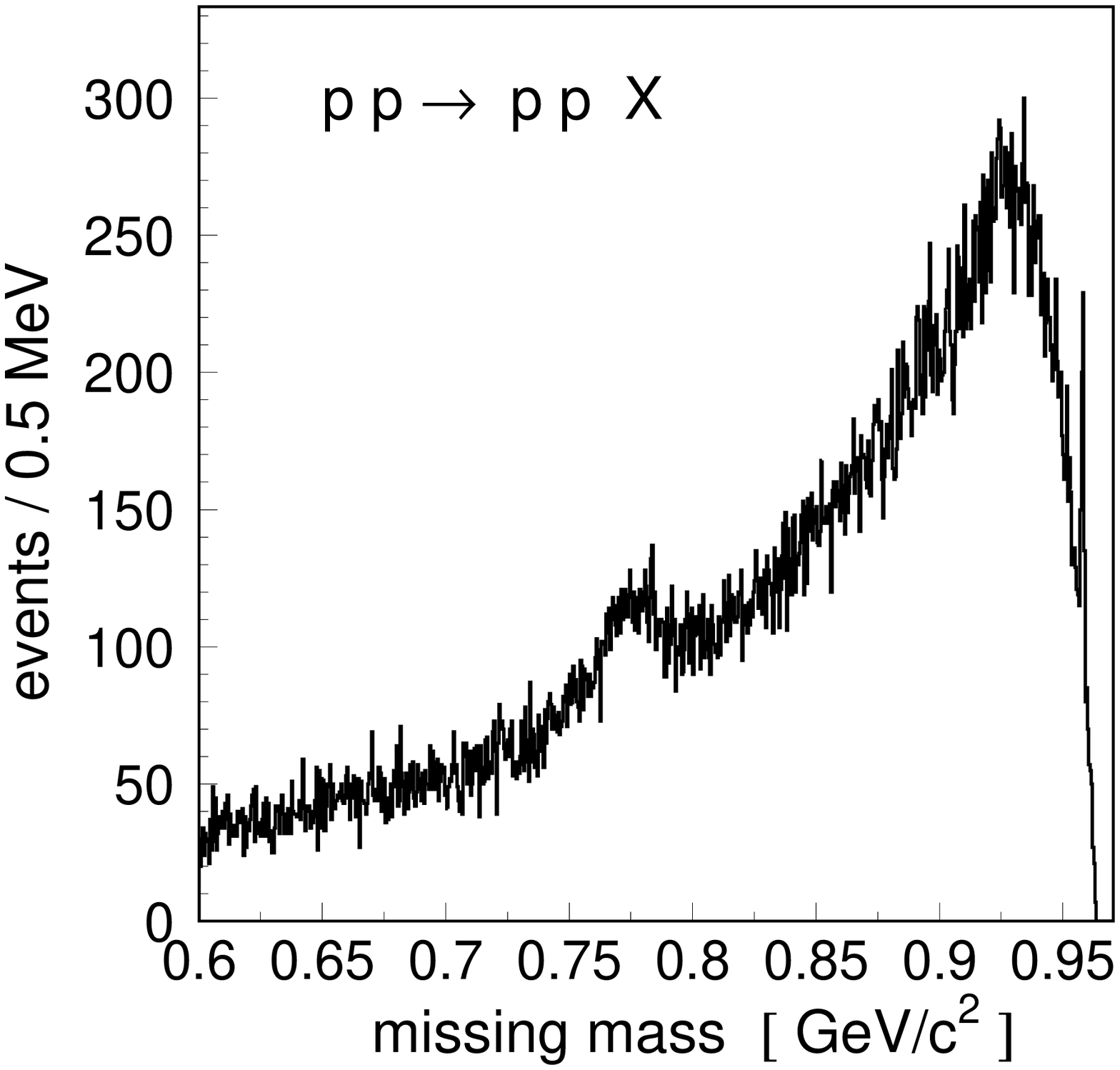,height=5.7cm,width=7.0cm,angle=0}
    }
       \put(1.0,13.8){
          {\large{ a)}}
       }
    \put(7.0,10.0){
       \epsfig{figure=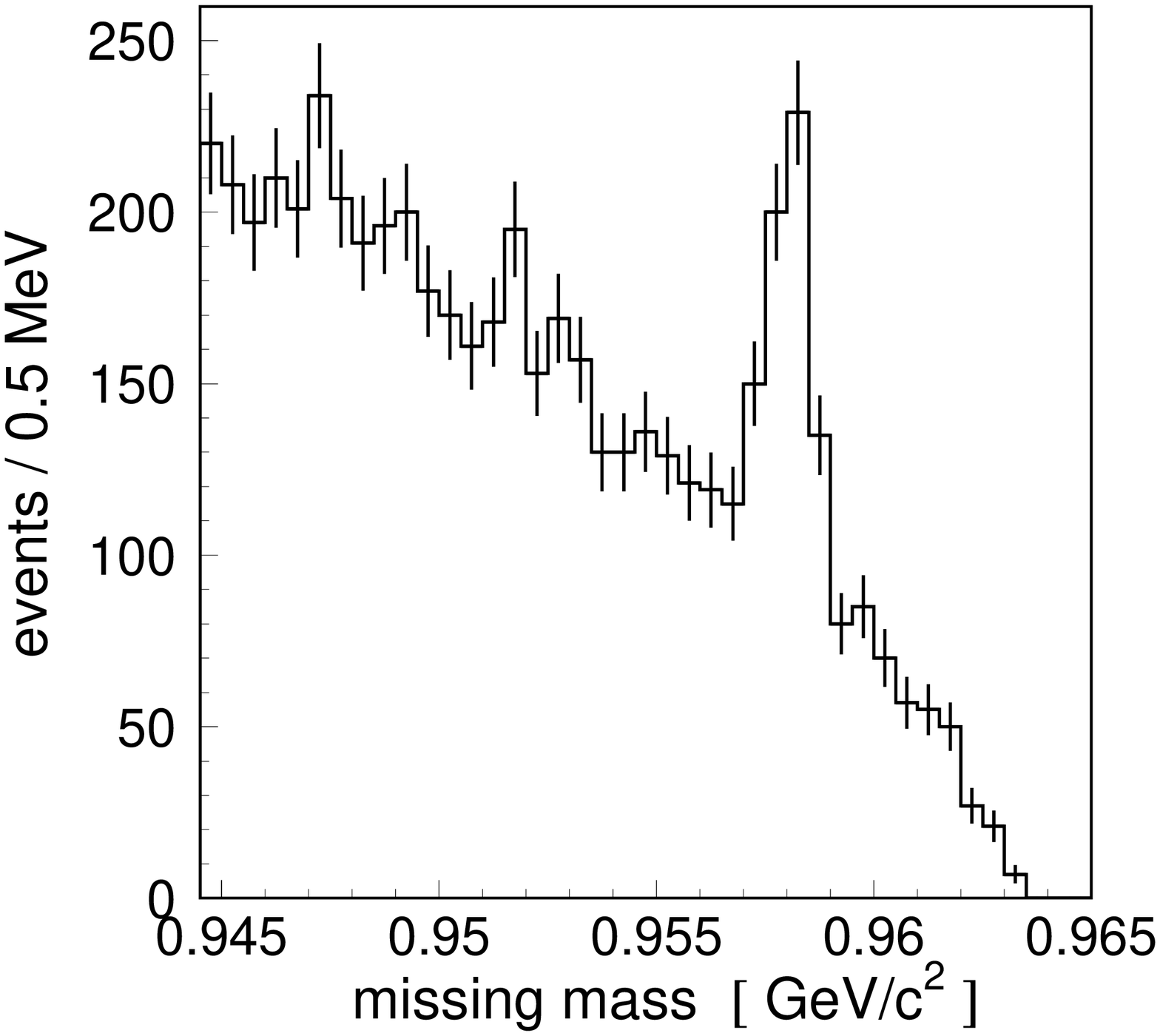,height=5.7cm,width=7.0cm,angle=0}
    }
       \put(12.5,13.8){
          {\large{ b)}}
       }
  \end{picture}
  \vspace{0.4cm}
  \caption{
           Missing mass of the unobserved particle or system of particles;
          {\bf upper row:}   measurements at Q~=~5.8~MeV above the $\eta^{\prime}$ production threshold; 
          {\bf middle row:}   at Q~=~7.6~MeV; and {\bf lower row:} at Q~=~1.5~MeV. 
          Background shown  as  dotted lines is combined from the measurements at different energies
          shifted to the appropriate kinematical limits and normalized to the solid-line histogram.
        }
\label{miss}
\end{figure}

 In the first step of the data analysis events with two tracks in drift chambers were preselected,
and the mass of each particle was evaluated. Figure~\ref{invmass} shows the squared mass of two
simultaneously detected particles. A clear separations is seen into groups of events with two protons,
two pions, proton and pion and also deuteron and pion. Thus, this spectrum allowed for a software 
selection of events with 
two registered protons. 
The knowledge of the momenta of both protons before and
after the reaction allows to calculate the mass of an unobserved particle or system of particles
created in the reaction. Figure~\ref{miss}a depicts the missing mass spectrum obtained for the
$pp \rightarrow pp X$ reaction at the excess energy value of Q~=~5.8~MeV above the $\eta^{\prime}$
meson production threshold. Most of the entries in this spectrum originate in the multi-pion production,
forming a continuous background to the well distinguishable peaks accounting for the $\omega$
and $\eta^{\prime}$ mesons production, which can be seen at  mass values of 782~MeV/c$^{2}$
and 958~MeV/c$^{2}$, respectively. 
The signal of the $pp\rightarrow pp\eta^{\prime}$ reaction
is better to be seen in the Figure~\ref{miss}b, where the missing mass distribution only in 
the vicinity of the kinematical limit is presented. Figure~\ref{miss}c shows the missing mass spectrum
for the measurement at Q~=~7.6~MeV together with the multi-pion background as 
combined  from the measurements at different excess energies~\cite{moskalnewrep}.
Subtraction of the background leads to the spectrum with a clear peak at the mass of the
meson $\eta^{\prime}$ as shown by the solid line in Figure~\ref{miss}d. 
The dashed histogram
in this figure corresponds to the Monte-Carlo simulations where the beam and target conditions
were deduced from the measurements of the elastically scattered protons~\cite{moskalnewrep}.
The scale of the simulated distribution was adjusted to fit the data, but the consistency of the 
widths is a measure of understanding of the detection system and the target-beam conditions. 
Histograms from a measurement at Q~=~1.5~MeV shown in Figures~\ref{miss}e,f demonstrate  
the achieved accuracy at the COSY-11 detection system.
The width of the missing mass distribution (Fig.~\ref{miss}f), which is now close to the natural width
of the $\eta^{\prime}$ meson ($\Gamma_{\eta^{\prime}}=0.203$~MeV~\cite{pdb98}), is again well reproduced by the
Monte-Carlo simulations. 

\section*{Results}
  \subsection*{Total cross section}
Determination of     (a) number of the produced $\eta^{\prime}$ events from the presented above missing mass 
distributions,   (b) luminosity from the simultaneous measurements of the elastically scattered protons,
and (c) detection system acceptance by means of the Monte-Carlo simulations 
allows for the calculations of the total cross section for the $pp\rightarrow pp\eta^{\prime}$ reaction.
The total cross section for the $pp\rightarrow pp\eta$ reaction was determined by the same method,
however, with a much bigger signal to background ratio (40/1) 
due to the larger total cross section values~\cite{smyrski}.    
\begin{center}
\vspace{-1.5cm}
\begin{figure}[H]
\psfig{figure=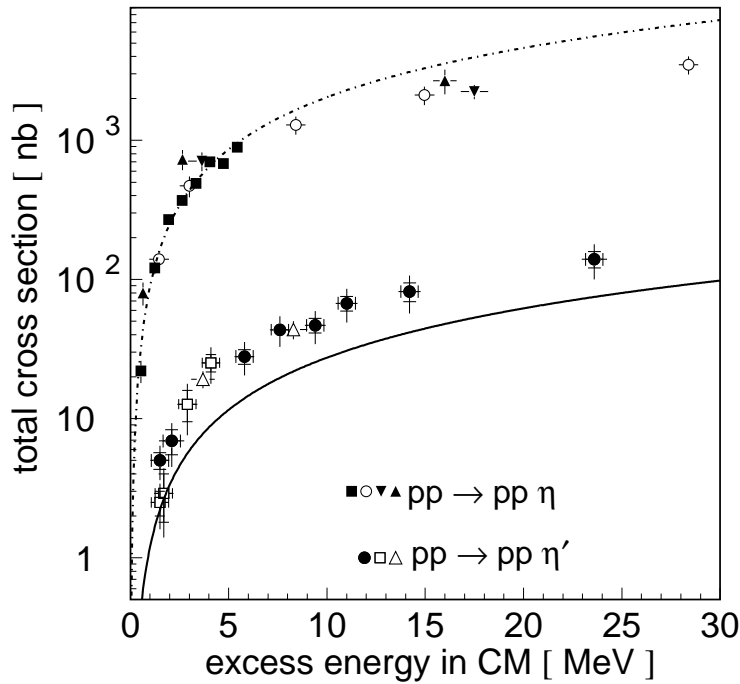,height=10.5cm,width=11.5cm,angle=0}
\caption{
          Total cross section for the reactions $pp\rightarrow pp\eta$ (upper points)
          and $pp\rightarrow pp\eta^{\prime}$ (lower points).
          Solid squares and solid circles corresponds to the yet unpublished COSY~-~11
          results. Triangles indicate measurements performed at SATURNE~\protect\cite{hiboupl,pinot},
          open circles at CELSIUS~\protect\cite{caleneta}, and open squares at COSY~\protect\cite{moskalprl}. 
          The curves are explained in the text.
        }
\label{etap-eta}
\end{figure}
\end{center}
 Figure~\ref{etap-eta} shows the compilation of the total cross sections for the $\eta$ and $\eta^{\prime}$
meson production together with the new COSY-11 data shown as filled squares~($\eta$) and filled
circles~($\eta^{\prime}$). The COSY-11 data on the $\eta$ production were taken  changing
continously during the measurement cycle
a momentum of the uncooled proton beam. 
This technique allowed for the precise
determination of the total cross section energy dependence near the kinematical threshold. The obtained
result confirmed the  enhancement of the close to threshold total cross section values
compared to the predictions
based on the phase space factors
and the proton-proton FSI which was earlier observed by  the PINOT~\cite{pinot}, WASA~\cite{caleneta},
 and SPES~III~\cite{hiboupl,bergdolt} collaborations.

\begin{center}
\vspace{-0.5cm}
\begin{figure}[H]
\epsfig{figure=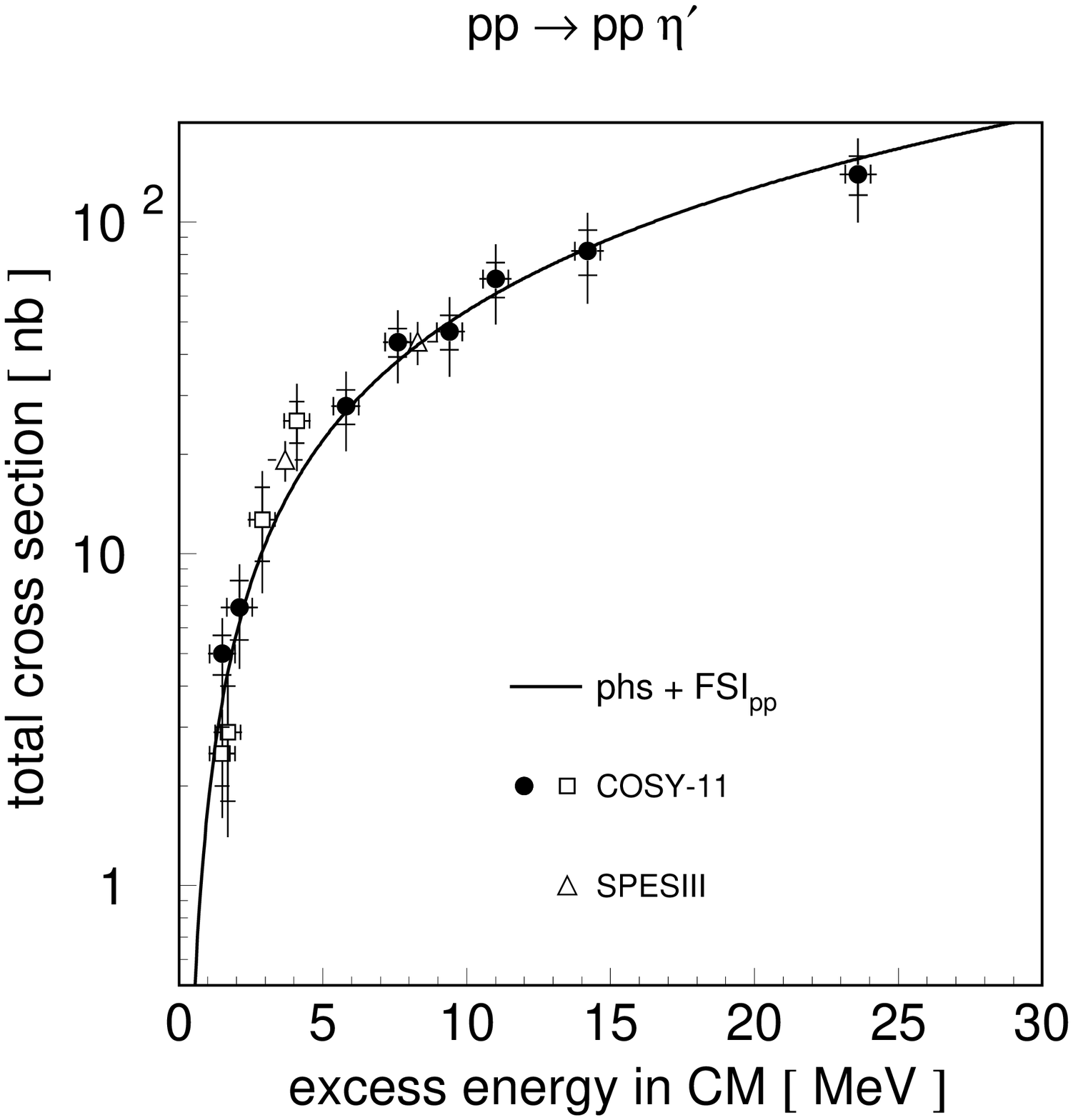,height=9.5cm,width=10.5cm,angle=0}
\caption{
         Total cross section for the $pp\rightarrow pp\eta^{\prime}$ reaction as a function 
         of the center of mass excess energy. 
         Open squares and triangles are from references~\protect\cite{moskalprl}
         and~\protect\cite{hiboupl}, respectively. Filled circles  indicate the results of the  analysis of the
         COSY~-~11 measurements performed in February 1998.
         The solid line depicts calculations of the total cross section under assumption that the
         primary production amplitude is constant and that only a proton-proton interaction take place
         in the exit channel. The proton-proton scattering amplitude was computed according to the formulas
         from reference~\protect\cite{druzhinin}. 
          The obtained energy dependence agrees within a few line thickness
         with the F{\"a}ldt and Wilkin model~\protect\cite{faldtwilk} presented in Figure~\ref{etap-eta}
         as a solid line. 
        }
\label{etap-fsi}
\end{figure}
\end{center}
New COSY~-~11 data concerning $\eta^{\prime}$ meson production 
are shown in Figure~\ref{etap-eta} as filled circles.
These measurements on the $pp \rightarrow pp\eta^{\prime}$ reaction were performed 
with the stochastically cooled proton beam and the integrated luminosity of 1360~nb$^{-1}$.
         Statistical and systematical errors are
         separated by dashes. The systematical error of the energy equals to 0.44~MeV
         constitutes of the 0.3~MeV due to the
         uncertainty in the detection system~\protect\cite{moskalabsolut} and 0.14~MeV due to the uncertainty
         in the $\eta^{\prime}$ meson mass~\protect\cite{pdb98}. The systematical error of the cross section values,
         including the overall normalization uncertainty,
         amounts to 15~$\%$~\protect\cite{moskalprl,moskalphd}.  
It is worth to stress again, that SPES~III and COSY~-~11 results obtained at different laboratories
are in a perfect agreement.

 The dashed-dotted line in  Figure~\ref{etap-eta}  
shows the energy dependence predicted by F\"aldt and Wilkin~\cite{faldtwilk}
normalized now to the COSY-11 data points,  and the solid line corresponds to the predictions based on a
one-pion-exchange model adjusted 
 to fit the close to threshold $pp\rightarrow pp\eta$ data (dashed line)~\cite{hiboupl}.
The  factor two discrepancy suggests that the short-range mechanisms may play a prominent role in the
production of these mesons~\cite{wilk,bass}.
 However, recent calculations performed by Nakayama et al.~\cite{nakayama} 
indicate that the determination of the total cross section 
close to threshold is surely not sufficient to establish
the contributions from different mechanisms to the overall production amplitude.
Specifically, the primary production amplitude for processes studied by these authors (Fig.~\ref{grafy}b,d,e) 
does not change significantly within the present experimental accuracy for the excess energies
below Q~=~30~MeV.  Therefore, 
the energy dependence of the total cross section for Q~$\leq$~30~MeV should be 
quite well described by the integral of the phase space volume weighted by the squared amplitude
of the final state interaction among the outgoing particles. And indeed, as shown in Figure~\ref{etap-fsi},
the data are in a good agreement with this model even 
without considering the $\eta^{\prime}$-proton interaction.
This leads to the conclusion that the $\eta^{\prime}$-proton interaction
is too weak to influence considerably, within the experimental error bars, the total 
cross section energy-dependence.

  \subsection*{Primary production amplitudes}
The cross section for the reaction $pp \rightarrow ppX$ can be expressed as:
 \begin{equation}
  \sigma_{pp\rightarrow ppX}=
  \frac{\displaystyle \int phase~space \cdot |M_{pp\rightarrow ppX}|^{2} }
       {flux~factor},
  \label{eqcrossform}
\end{equation}
where, $M_{pp\rightarrow ppX}$ denotes the transition matrix element
for the $pp\rightarrow ppX$ reaction, and X stands for $\pi^{0},\eta$ or $\eta^{\prime}$ mesons.
In analogy with the {\em Watson-Migdal} approximation~\cite{wats52} for two body processes,
it can be assumed that the complete transition amplitude of a production process
$M_{pp\rightarrow ppX}$ factorizes approximately as~\cite{moalem1}:
\begin{equation}
   M_{pp \rightarrow ppX  } \approx  M_{0} \cdot M_{FSI}
\end{equation}
where,  $M_{0}$ accounts for  all possible
production processes, and $M_{FSI}$  describes the elastic interaction of protons
and X meson in the exit channel.
Making further assumptions that  only the proton-proton interaction is present in the exit channel
(\begin{math} M_{FSI} = M_{pp\rightarrow pp} \end{math})
and that the primary production amplitude does not change with the excess energy,
it is possible  to calculate $|M_{0}|$.
The enhancement from the 
proton-proton interaction, $|M_{pp\rightarrow pp}|^{2}$,
was estimated as an inverse of the squared Jost function,
with Coulomb interaction being taken into account~\cite{druzhinin}.
The $|M_{pp\rightarrow pp}|^{2}$ is a dimensionless factor which
turns to zero with vanishing relative
protons momentum k, peaks sharply at k$\approx$25~MeV/c and 
approaches asymptotically unity for large proton-proton relative momenta.
\begin{center}
\begin{figure}[H]
 \unitlength 1.0cm
  \begin{picture}(12.2,12.0)
    \put(-0.5,0.0){
       \epsfig{figure=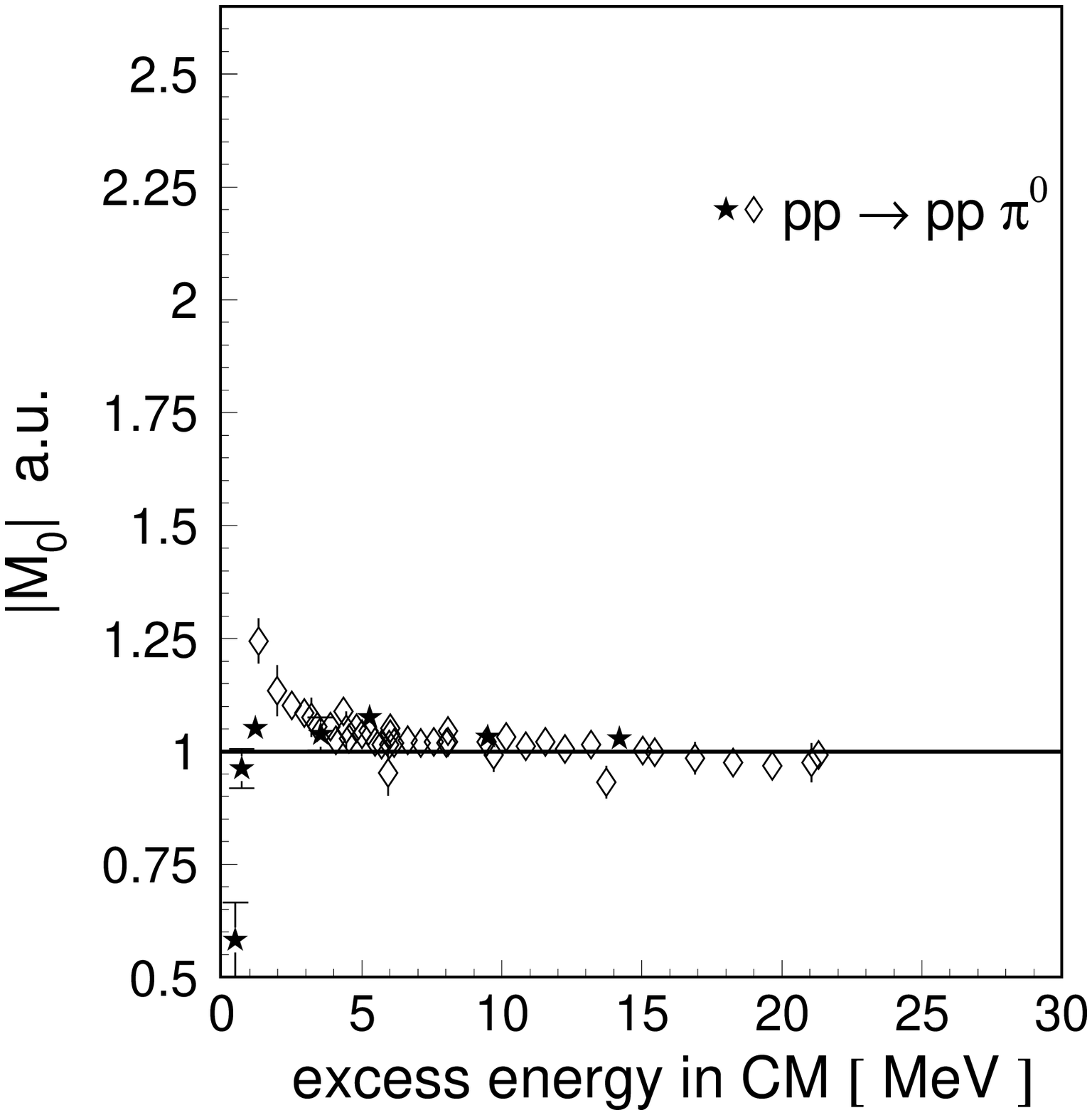,height=6.5cm,width=7.2cm,angle=0}
    }
    \put(6.8,0.0){
       \epsfig{figure=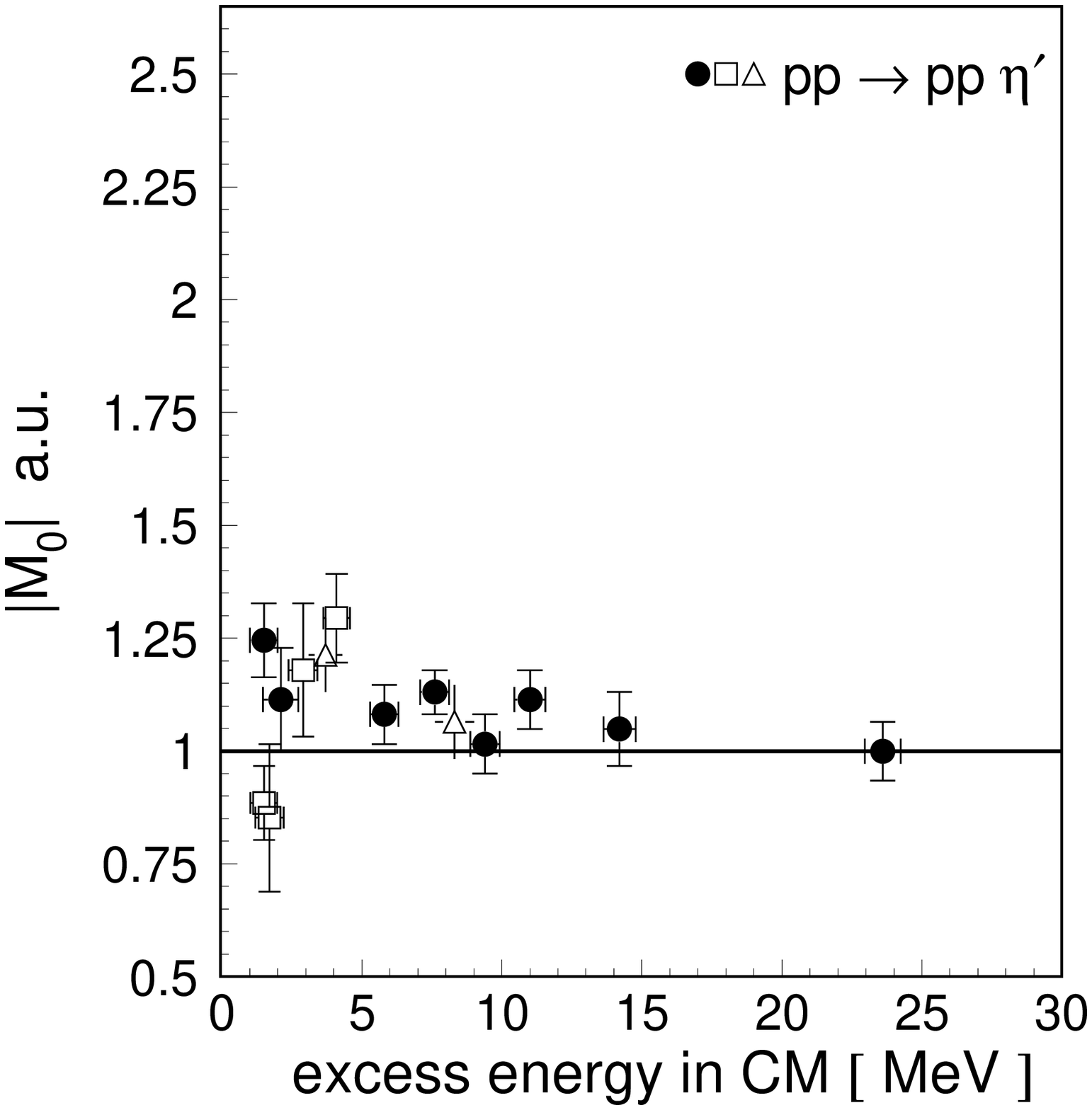,height=6.5cm,width=7.2cm,angle=0}
    }
    \put(3.0,6.2){
       \epsfig{figure=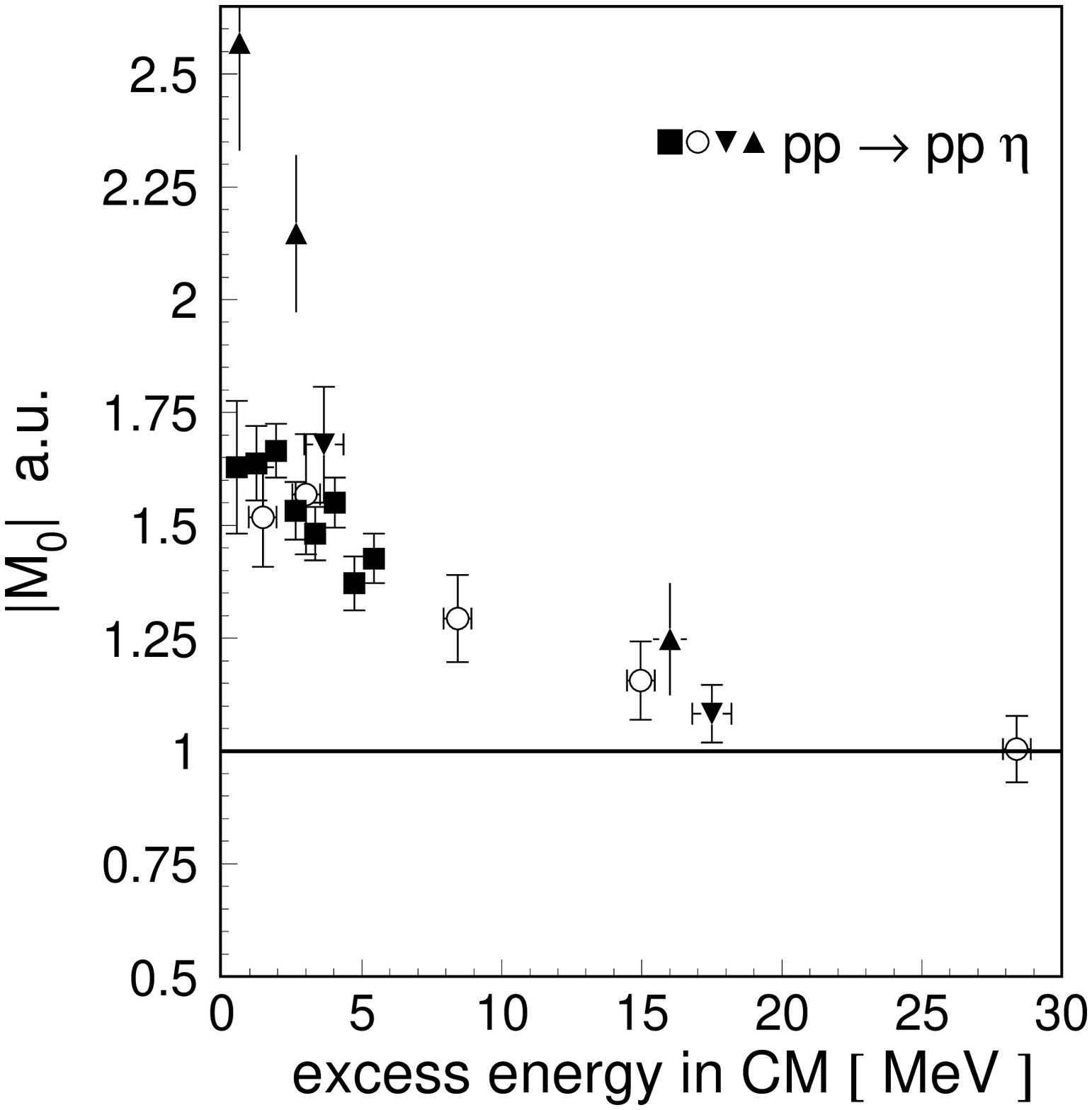,height=6.5cm,width=7.2cm,angle=0}
    }
  \end{picture}
  \vspace{0.4cm}
  \caption{
          Quantity $|M_{0}|$ extracted from the experimental data
          for the reactions  $pp \rightarrow pp\eta$ --- upper picture; \ \  \
          $pp \rightarrow pp\pi^{0}$ --- left picture; \ \ \ 
          $pp \rightarrow pp\eta^{\prime}$ --- right picture; 
        }
\label{m0}
\end{figure}
\end{center}

Figure~\ref{m0} compares the extracted
 absolute values for the modulus of the primary production amplitude for the near-threshold production
of the $\eta$, $\pi^{0}$ and $\eta^{\prime}$  mesons.
The quantity $|M_{0}|$ is normalized to unity
at the point of highest excess energy, for each meson separately. 
If the performed assumptions in the derivation of $|M_{0}|$ were fulfilled
the obtained values would be equal to one as depicted by the solid line.
It can be seen, however, that in the case of the $\eta$ meson, $|M_{0}|$ grows 
with decreasing excess energy reflecting  attractive $\eta$-proton interaction. 
In the data for the $\pi^{0}$ production, 
   apart from the two closest-to-threshold points\footnote{
   Due to the steep falling of the total cross section near-threshold 
   already a small change of the energy (0.2~MeV) lifts the points significantly up.
   Moreover, for the very low energies nuclear and Coulomb scattering are expected to compete.
   The limit is  approximately at 0.8~MeV of the proton energy in the rest system of the other proton,
   where the Coulomb penetration factor~C$^{2}$ is equal to 0.5~\cite{jackson}.
   Thus, one should be careful, at small excess energies,
   where the approximately treated Coulomb interaction dominates.
  },
one can notice a tiny grow of $|M_{0}|$ when the excess energy decreases from Q~=~20~MeV to Q~=~2~MeV.
This may be cause by the small $\pi$-proton interaction. The deviation from the constant is much smaller
than in the $\eta$ meson case since, the S-wave $\pi$-proton interaction is much weaker than the $\eta$-proton
one. 

Similarly, neglecting the two lowest points for the  $\eta^{\prime}$ meson, one observes about 20~$\%$ increase
of $|M_{0}|$ when approaching the threshold. 
This may indicate a small attractive $\eta^{\prime}$-proton interaction. Anyhow, with the new COSY~-~11 data points
the possible $\eta^{\prime}$-proton repulsive interaction must be excluded.

Instead of conclusion 
the article of Bernard et al.~\cite{bernard} is recommended
where the threshold matrix-elements for the $pp\rightarrow pp\pi^{0}(\eta,\eta^{\prime})$ reactions were evaluated
in a fully relativistic Feynman diagrammatic approach as reported by N.~Kaiser at this conference.

\section*{Acknowledgments}
    P.M. acknowledges the hospitality and financial support from the Forschungszentrum 
    J{\"u}lich and the  Foundation for Polish Science.

\end{document}